\documentclass[superscriptaddress,twocolumn]{revtex4}
\usepackage{amssymb}
\usepackage{amsfonts}
\usepackage{amsmath}
\usepackage{graphicx}
\usepackage{dcolumn}
\usepackage{bm}
\usepackage[latin1]{inputenc}
\usepackage[spanish]{babel}
\usepackage[dvips]{hyperref}

\begin{document}
\title{Nonsingular Promises from Born-Infeld Gravity}

\author{Franco Fiorini}
\email{francof@cab.cnea.gov.ar}\thanks{Member of Carrera del Investigador Cient\'{\i}fico (CONICET,
Argentina)}  \affiliation{Centro Atómico
Bariloche, Comisión Nacional de Energía Atómica R8402AGP
Bariloche, Argentina.}


\begin{abstract}
Born-Infeld determinantal gravity formulated in Weitzenb\"{o}ck spacetime is discussed in the context of Friedmann-Robertson-Walker (FRW) cosmologies. It is shown how the standard model big bang singularity is absent in certain spatially flat FRW spacetimes, where the high energy regime is characterized by a de Sitter inflationary stage of geometrical character, i.e., without the presence of the inflaton field. This taming of the initial singularity is also achieved for some spatially curved FRW manifolds where the singularity is replaced by a de Sitter stage or a big bounce of the scale factor depending on certain combinations of free parameters appearing in the action. Unlike other Born-Infeld-like theories in vogue, the one here presented is also capable of deforming vacuum general relativistic solutions.
\end{abstract}

\maketitle

After the proposal addressed in \cite{Deser}, the quest for nonsingular classical
spacetimes was carried out several times in the context of extended theories
of gravity with Born-Infeld-like structure \cite{Fein1}-\cite%
{Vollick}. The choice of Born-Infeld actions for the gravitational field is
strongly suggested for taming the singularities present in Einstein's
theory, especially if one remembers the early success of the theory in the
analogue framework of classical electrodynamics, i.e., in the
problem concerning the infiniteness of the pointlike charge self-energy
characterizing Maxwell's theory \cite{BornI}. Remarkable enough, the Born-Infeld
electromagnetic action also plays a prominent role in string theory, being
the proper action for describing the electromagnetic field in $D$-branes \cite%
{Tsey2}, \cite{Tsey1}, and it was shown recently that Born-Infeld gravitational structures of the sort considered in references \cite{Tek1} and \cite{Tek2} appear naturally as counterterms in four-dimensional anti-de Sitter spacetime \cite{jatkar}. However, in all the above mentioned gravitational
Born-Infeld schemes, and due to the fourth order field equations for the
metric tensor coming from the higher order curvature components in the
action, no single \emph{exact} solution describing a regular spacetime has ever been
found.

Along this quest, another closely related nonsingular programme for the gravitational field,
the so-called Eddington inspired Born-Infeld gravity, was introduced by Ba%
\~{n}ados in \cite{Max1} and studied carefully very recently, mostly in
cosmological and astrophysical environments \cite{Max2}-\cite{Sham}. Second order motion equations in this framework are assured due
to the independent role played by the metric and the connection, and a number of exact cosmological
solutions without the big bang singularity were found \cite{Max3}, \cite%
{Avelino2}, even though the theory seems to also possess singular states \cite{pani}, \cite{chen}. Nevertheless, Eddington-Born-Infeld gravity differs from
Einstein theory only when matter sources are present, an unfortunate fact
that deprives us from regular black hole states in pure vacuum.

The purpose of this work is to present a number of regular cosmological solutions arising from the previously featured
Born-Infeld determinantal gravity \cite{nos4}. In many senses, this theory
captures the most distinguished points of all the above mentioned Born-Infeld-like schemes, for it assures second order motion equations for the vielbein
field $e^{a}$ and, unlike Eddington-Born-Infeld gravity, it is also capable of
deforming vacuum general relativistic spacetimes. Actually, in Ref. \cite%
{nos4} we have shown how the conical singularity present in vacuum three-dimensional Einstein gravity was deformed into a geodesically complete
curved spacetime by solving exactly the gravitational Born-Infeld motion
equations.

\bigskip

Born-Infeld-like actions are not only good candidates in order to deal with singularities, but they are also quite natural. By its very nature, the Lagrangian density $L$ must be a scalar density of weight one and, in a $D$-dimensional orientable manifold, the Lagrangian $\widetilde{L}$ is represented by an $D$-form which locally looks like $\widetilde{L}=L(\phi,\partial^j\phi) dx^1\wedge...\wedge dx^n$. Here we suppose that $L$ depends on certain fields $\phi$ and its derivatives up to order $j$. In order to construct this density we have at hand a canonical procedure; just take a linear combination of squared roots of determinants of second rank tensors. Hence we can write in general
\begin{equation}
L(\phi,\partial^j\phi)=\alpha_{k}\sum_{k}\sqrt{|L_{\mu\nu}^{(k)}|},
\label{combinacion}
\end{equation}
where $|\;|$ stands for the absolute value of the determinant. Formally, each one of the tensors in the combination (\ref{combinacion}) can be decomposed according to
\begin{equation}
L_{\mu\nu}=\lambda_{0}g_{\mu\nu}(\phi)+\lambda_{1}F_{\mu
\nu}^{(1)}(\phi,\partial \phi)+...+\lambda_{j}F_{\mu
\nu}^{(j)}(\phi,...,\partial^{j}\phi), \label{spliting}
\end{equation}
where $\lambda_{i}$, $0\leq  i\leq j$, are arbitrary couplings at this point. The splitting (\ref{spliting})
emphasizes that $L_{\mu\nu}$ can be viewed formally as a sum of tensors $F_{\mu
\nu}^{(i)}(\phi,...,\partial^{i}\phi)$ containing derivatives up to order $i$, where we have written $F_{\mu
\nu}^{(0)}(\phi)\equiv g_{\mu\nu}(\phi)$. This last tensor has a prominent role in (\ref{spliting}) because it contains no derivatives at all.

Even though we do not account with a general rule for the number of terms we have to consider in (\ref{spliting}), a sufficient condition in order to obtain second order differential equations for the fields $\phi$ results by interrupting the sum in $F_{\mu\nu}^{(1)}(\phi,\partial \phi)\equiv F_{\mu \nu}$ \footnote{It could be the case that specific combinations of tensors $F_{\mu\nu}$ containing up to second derivatives of the fields could lead to second order motion equations too, as, for instance, in Lovelock actions. We will not deal with those (important) subtleties in this work.}. Additionally, this condition also restricts the summands in (\ref{combinacion}) to three, otherwise we would find redundant terms in the Lagrangian. So, the Lagrangian takes the form
\begin{equation}
L=\alpha_{0}
 \sqrt{|g_{\mu\nu}|}+\alpha_{1}\sqrt{|g_{\mu\nu}+2\lambda^{-1} F_{\mu \nu}|}+\alpha_{2}\sqrt{| F_{\mu \nu}|},
\label{tres}
\end{equation}
where we called $2\lambda^{-1}\equiv\lambda_{1}/\lambda_{0}$, and a redefinition of the constants $\alpha_{k}$ ($k=0,1,2$) was performed. An additional reduction will assure the proper low field limit of the theory; taking $\alpha_{0}=-\alpha_{1}$ and $\alpha_{2}=0$ the Lagrangian will depend on just one undetermined constant (say $\lambda$), and the action will finally acquire the form
\begin{equation}
I_{\mathbf{BIG}}=\frac{\lambda}{16 \pi G} \int d^{D}x\Big[\sqrt{|g_{\mu
\nu}+2\lambda ^{-1}F_{\mu \nu }|}-\sqrt{|g_{\mu \nu }|}\Big].
\label{acciondetelectro}
\end{equation}
Note that, whatever the (at this point unspecified) tensor $F_{\mu \nu }$ is, the low energy limit ($\lambda
\rightarrow \infty $) of the action (\ref{acciondetelectro}) can be easily
obtained by factoring out $\sqrt{|g_{\mu \nu }|}$ and using
\begin{equation}
\sqrt{|\mathbb{I}+2\lambda ^{-1}\mathbb{F}|}=1+\lambda ^{-1}Tr(\mathbb{F})+%
\mathcal{O}(\lambda ^{-2}),  \label{primer}
\end{equation}%
where $\mathbb{F}\equiv F_{\mu }^{\,\,\nu }$ and $\mathbb{I}$ is the
identity. Hence, in the low energy limit we get the action
\begin{equation}
I_{\downarrow}=\frac{1}{16\pi G}\int d^{D}x\,\sqrt{|g_{\mu \nu }|}%
\,Tr(\mathbb{F})\;.  \label{accionbaja}
\end{equation}%
It only remains to find the tensors $g_{\mu\nu}$ and $F_{\mu \nu }$. Note that $F_{\mu \nu }$ contains just first derivatives of the dynamical field $\phi$, so to look for it in a Riemannian context would be futile; no second rank tensor built with first derivatives of the metric will give rise through its trace to the scalar curvature $R$ characterizing the Hilbert-Einstein action of general relativity (GR).

To link this low energy action to the Einstein-Hilbert action, we can evoke
the teleparallel representation of GR. The linkage between the standard (Riemannian) description of GR and its absolute parallelism (Weitzenb\"{o}ck) version is summarized in the equation
\begin{equation}
T\,=\,-R+2\;e^{-1}\;\partial _{\nu }(e\,T_{\sigma }^{\ \sigma \nu }\,)\;,
\label{divergence}
\end{equation}%
where $T$ is the so-called Weitzenb\"{o}ck invariant,
\begin{equation}
T\ \equiv S_{\rho }^{\;\mu \nu }\ T_{\;\mu \nu }^{\rho }\,,
\label{Weitinvar}
\end{equation}%
with $S_{\rho }^{\;\;\mu \nu }$ defined as
\begin{equation}
S_{\rho }^{\;\;\mu \nu }\equiv \frac{1}{4}\,(T_{\rho }^{\;\mu \nu
}-T_{\;\;\;\rho }^{\mu \nu }+T_{\;\;\;\rho }^{\nu \mu })+\frac{1}{2}\ \delta
_{\rho }^{\nu }\ T_{\sigma }^{\ \ \sigma \mu }-\frac{1}{2}\ \delta _{\rho
}^{\mu }\ T_{\sigma }^{\ \ \sigma \nu },  \label{tensor}
\end{equation}%
and $T_{\rho }^{\,\,\,\mu \nu }$ are the components of the Weitzenb\"{o}ck
torsion $T^{a}\equiv de^{a}$, i.e., $T_{\;\mu \nu }^{\rho }=e_{a}^{\rho }\,(\partial _{\mu }e_{\nu }^{a}-\partial
_{\nu }e_{\mu }^{a})$. This torsion emerges out from Weitzenb\"{o}%
ck connection $\Gamma _{\mu \nu }^{\rho }=\,e_{a}^{\rho }\,\partial _{\nu
}e_{\mu }^{a}$ (here $e_{a}^{\rho }$ is the inverse of $e_{\mu }^{a}$, so $%
e_{a}^{\rho }e_{\rho }^{b}=\delta _{a}^{b}$). In the Weitzenb\"{o}ck representation of GR, the dynamical field is the vielbein $e^a$, and the metric is a subsidiary field related to the vielbein in the form
\begin{equation}
g_{\mu \nu }=\eta _{ab}e_{\,\,\mu }^{a}e_{\,\,\nu }^{b}.  \label{metric}
\end{equation}%
This last equation determines the tensor $g_{\mu \nu }$ in terms of the fundamental fields $\phi=e^a$.
This implies $e\equiv \det (e_{\,\,\mu }^{a})=\sqrt{|g_{\mu \nu }|}$, which appears in (\ref{divergence}). According to Eq. (\ref{divergence}), the theories described by the Lagrangian densities $e\,T$ and $e\,R$ are dynamically equivalent, because these two quantities differ in a total derivative. Therefore,
the low energy regime governed by action (\ref{accionbaja}) will be GR provided $Tr(\mathbb{F})=T$. This is the sole constraint for the
tensor $F_{\mu \nu }$ in the Born-Infeld gravitational action (\ref{acciondetelectro}).

We are now in position to find $F_{\mu \nu }$. A direct inspection shows that the more general candidate should read
\begin{equation}
F_{\mu \nu }=\alpha \,S_{\mu }^{\;\;\lambda \rho }T_{\nu \lambda \rho
}+\beta \,S_{\lambda \mu }^{\;\;\;\rho }T_{\,\,\,\,\,\nu \rho }^{\lambda
}+\gamma \,g_{\mu \nu }\,T,  \label{tensorF}
\end{equation}%
where $\alpha ,\beta ,\gamma $ are dimensionless constants such as $\alpha
+\beta +D\, \gamma =1$, hence, ensuring that $Tr(\mathbb{F})=T$ \footnote{In the original construction of Ref. \cite{nos4} we have considered the theory with $\gamma=0$. I give my thanks to F. Canfora for aware me about the inclusion of a pure trace term.}. With the definitions (\ref{metric}) and (\ref{tensorF}), the action (\ref{acciondetelectro}) is now determined. It governs the dynamics of the vielbein field $e^a$ and contains just first derivatives of it, thus guaranteing second order differential equations. It also assures that in regions where $\mathbb{F}\ll\lambda$, the gravitational phenomena are those predicted by Einstein's theory. Note that the particular choice $\alpha=\beta=0$ trivializes the determinantal character of action (\ref{acciondetelectro}), for in this case we actually have that the action becomes
\begin{equation}
I_{\mathbf{BI0}}=\frac{\lambda}{16 \pi G} \int d^{D}x\,\sqrt{|g_{\mu
\nu}|}\,\Big[\Big(1+2\frac{T}{D\lambda}\Big)^{D/2}-1\Big],
\label{accionbio}
\end{equation}
which is an $f(T)$-type action. In fact this constitutes the sole $f(T)$-like action that can be obtained from the determinantal structure (\ref{acciondetelectro}). Finally, we should mention an important point. The tensor (\ref{tensorF}) is not invariant under local Lorentz transformations of the vielbein $e^a$, so neither is the action (\ref{acciondetelectro}). However, it is mandatory to note that the breaking of Lorentz invariance happens at a Born-Infeld scale of order $\ell_{BI}^2=\lambda^{-1}$. This is an exceedingly small length scale possibly associated with $\ell_{p}$, the Planck length, where no fully satisfactory description of the spacetime structure seems to exist currently.

\bigskip
In order to exhibit the power hidden in action (\ref{acciondetelectro}), from now on we will focus on cosmological scenarios. For this reason we will be conservative and fix $D=4$. It is our intention to deal first with spatially flat FRW-like spacetimes. For this purpose we have the frame $e^a=diag(1,a(t),a(t),a(t))$, which leads to the spatially flat FRW line element
\begin{equation}
ds^2=dt^2-a^2(t)[dx^2+dy^2+dz^2], \label{valink=0}
\end{equation}%
 with $a(t)$ the scale factor. Again, being very conservative we shall consider as a source a perfect fluid with state equation $p=\omega \rho$, where $\omega$ is the barotropic index. In this case we can easily obtain the initial value equation, i.e., the motion equation coming from varying the action with respect to $e^0_{\,\,0}$. By using the comoving frame, the energy-momentum simply reads $T^\mu_\nu=diag(\rho,-\omega\rho,-\omega\rho,-\omega\rho)$, and the initial value equation results
\begin{equation}
\frac{\sqrt{1-B H^2}}{\sqrt{1-A H^2}}[1+2B H^2-3 AB H^4]-1=\frac{16 \pi G }{\lambda}\rho, \label{valink=0gen}
\end{equation}%
where
\begin{equation}
A=6(\beta+2\gamma)/\lambda,\,\,\,\,\,\,B=2(2\alpha+\beta+6\gamma)/\lambda. \label{condiciones}
\end{equation}
This is the modified Friedmann equation governing the dynamics of the Hubble rate $H=\dot{a}/a$. The energy density $\rho$ is linked to the scale factor by the conservation equation $\dot{\rho}+3(\rho+p)H=0$. If one is dealing with a perfect fluid state equation this conservation law assumes the form
\begin{equation}
\rho(t)=\rho_{0}\Big(\frac{a_{0}}{a(t)}\Big)^{3(1+\omega)}, \label{conservacion}
\end{equation}%
where $\rho_{0}$ and $a_{0}$ are two constants alluding to present-day values. If we replace (\ref{conservacion}) in the right-hand side of (\ref{valink=0gen}), we get a first order differential equation for the scale factor $a(t)$. The other motion equations coming from the remaining tetrad components $e^a_{\mu}$, if non-null, are just a consequence of combining the time derivative of (\ref{valink=0gen}) with (\ref{conservacion}), so they do not provide additional information.

It is then convenient to carefully inspect the equation (\ref{valink=0gen}) for a case that, due to its simplicity, is of particular interest, namely, when $B=0$. In addition to $B=0$, the normalization condition $\alpha+\beta+4\gamma$ leads us to $A=12/\lambda$, so the motion equation reduces to
\begin{equation}
\frac{1}{\sqrt{1-\frac{12H^2}{\lambda}}}-1=\frac{16 \pi G }{\lambda}\rho. \label{BI0}
\end{equation}%
Surprisingly enough \footnote{It is important to be aware that the theory considered here and the one of Refs. \cite{Nos} and \cite{Nos2} are totally different. The equivalence is valid at the level of this particular solution.}, Eq. (\ref{BI0}) is the same as the obtained previously in the Born-Infeld gravitational theory with $f(T)$ structure discussed in Refs. \cite{Nos} and \cite{Nos2}. This field equation conduces to an exact solution with remarkable properties which we just summarize here (see the mentioned references for details). For every barotropic index $\omega>-1$ (radiation and dust matter lying in this interval) the scale factor describes a regular (geodesically complete) spacetime without the big bang singularity and possessing a natural inflationary stage of geometrical character. Actually it can be easily seen that
\begin{equation}
a(t\rightarrow -\infty) \propto exp\Big[\sqrt{\frac{\lambda}{12}}\,t\Big],\label{inflation1}
\end{equation}%
so there exists a maximum Hubble factor $H_{max}=\sqrt{\lambda/12}$ as we backtrack the cosmic evolution. This maximum Hubble factor assures for the early Universe a $\lambda$-driven de Sitter evolution of infinite duration, and it is responsible for the geodesic completeness of the spacetime, for every past directed timelike or null geodesic can be extended to arbitrary values of the affine parameter. As a consequence of the presence of $H_{max}$, it can be easily shown that the invariants of the geometry also reach saturation values given by $R_{max}=\lambda$, $(R^{\mu\nu}R_{\mu\nu})_{max}=\lambda^2/4$ and $(R^{\mu}_{\,\,\,\nu\rho\sigma}R_{\mu}^{\,\,\,\nu\rho\sigma})_{max}=\lambda^2/6$ as $t\rightarrow -\infty$.

\bigskip

The characterization of isotropic and homogeneous cosmological manifolds would not be complete if we would cease the analysis in the spatially flat FRW models. In order to carry on with the investigation let us deal now with spatially curved cosmological models. The frames adapted for this symmetry are substantially more complicated than the simple diagonal frames corresponding to the spatially flat case worked above. In Ref. \cite{Nos3} it was showed that a global basis of frames for spatially curved FRW spacetimes reads
\begin{eqnarray}
e^0&=&dt,\notag\\
e^1&=&a(t)\, E^1, \notag\\
e^2&=&a(t)\, E^2,\notag \\
e^3&=&a(t)\, E^3,
\label{tetcurv}
\end{eqnarray}
where the 1-forms $E^1$, $E^2$, and $E^3$ are
\begin{eqnarray}
\frac{E^1}{k}&=&- k \cos\theta \,d\psi+ \sin(k\psi)\sin\theta\,\cos(k\psi)\, d\theta- \notag\\
&& -\sin^2(k\psi)\sin^2\theta \,d\phi,\notag \\
\frac{E^2}{k}&=&\,\,\,\,k \sin\theta\cos\phi \,d\psi - \notag\\
&&- \sin^2(k\psi)[\sin\phi-\cot(k\psi)\cos\theta\cos\phi]\,d\theta-\notag\\
&&-\sin^2(k\psi)\sin\theta[\cot(k\psi)\sin\phi+\cos\theta\cos\phi]\,d\phi,\notag\\
\frac{E^3}{k}&=&-k \sin\theta\sin\phi \,d\psi - \notag\\
&&- \sin^2(k\psi)[\cos\phi+\cot(k\psi)\cos\theta\sin\phi]\,d\theta-\notag\\
&&-\sin^2(k\psi)\sin\theta[\cot(k\psi)\cos\phi-\cos\theta\cos\phi]\,d\phi.\notag\\
\label{campos}
\end{eqnarray}
This global basis leads to the line element
\begin{equation}
ds^2=dt^2-a^2(t)k^2[d(k\psi)^2-\sin^2(k\psi)(d\theta^2+\sin^2\theta \,d\phi^2)], \label{metcurv}
\end{equation}
where $(\psi,\theta,\phi)$ are standard spherical coordinates. The parameter $k$ appearing in Eqs. (\ref{campos}) and (\ref{metcurv}) takes the values $k=1$ for the spatially spherical Universe and $k=i$ for the spatially hyperbolic one.

The equations of motion taking into account the full parameter space $(\alpha,\beta,\gamma)$ are indeed very complicated for the spatially curved manifolds under consideration, and it is not our intention to deal with them in their full generality. In analogy with the study made before, we shall focus on a few cases of particular interest for our present concerns. The fine solution encoded in Eq. (\ref{BI0}) has an equally nice counterpart in the curved case. Actually, the motion equation for $B=0$ reads
\begin{equation}
\frac{(1\pm \frac{1}{\lambda a^2})^{3/2}}{\sqrt{1-\frac{12 H^2}{\lambda}}}-1=\frac{16 \pi G }{\lambda}\rho, \label{curvcaso1}
\end{equation}
where from now on $+$ and $-$ correspond to the closed and open cases, respectively. An exact solution of this equation is rather elusive, but we can extract its most important features by writing it in the form
\begin{equation}
\dot{y}^2+V(y)=0, \,\,\,\,\,\,y=\frac{a(t)}{a_{0}} \label{ecpot}
\end{equation}%
after defining the \emph{effective potential} $V(y)$ given by
\begin{equation}
V(y)=-\frac{\lambda}{12}y^2\Big[1-\frac{(1\pm k_{0}\, y^{-2})^3}{(1+\beta_{0}\,y^{-3(1+\omega)})^2}\Big],\label{potcaso1}
\end{equation}%
where we have defined the constants $k_{0}=1/\lambda a_{0}^2$ and $\beta_{0}=16 \pi G \rho_{0}/\lambda$.

If we focus on the high energy regime where the theory supposes to makes a difference with respect to the GR singular behavior, we can expand the potential (\ref{potcaso1}) in the small quantity $y=a(t)/a_{0}$ (from now on we will take $\omega=1/3$). Under this circumstance, Eq. (\ref{ecpot}) results in
\begin{equation}
\dot{y}^2-\frac{\lambda}{12}y^2=\mathcal{O}(y^4)\approx0, \label{ecpotbajo}
\end{equation}%
Naturally, we have then
\begin{equation}
a(t)\approx exp\Big[\sqrt{\frac{\lambda}{12}}\,t\Big],\,\,\,as\,\,\, a(t)/a_{0}\rightarrow 0.  \label{solu}
\end{equation}%
Again, as in its flat counterpart of Eq. (\ref{inflation1}), we see how the initial singularity is removed by the presence of an inflationary early stage driven by the Born-Infeld constant $\lambda$. Note that this result is independent of the open or closed character of the spatial slices.

\bigskip

Some other interesting results emerge when one considers the case $\beta=0$, $\alpha-12\gamma=0$. For this particular choice of parameters the initial value motion equation reads
\begin{equation}
\frac{1\pm \frac{1}{\lambda a^2}}{\sqrt{1\pm \frac{1}{\lambda a^2}-\frac{12 H^2}{\lambda}}}-1=\frac{16 \pi G }{\lambda}\rho. \label{curvcaso2}
\end{equation}%
This equation would be even harder to solve than the one of the previous case (Eq. (\ref{curvcaso1})), so again we will put it in the form (\ref{ecpot}). Now the effective potential results
\begin{equation}
V(y)=-\frac{\lambda}{12}y^2\Big[1\pm k_{0}\,y^{-2}-\Big(\frac{1\pm k_{0}\, y^{-2}}{1+\beta_{0}\,y^{-3(1+\omega)}}\Big)\Big].\label{potcaso2}
\end{equation}%
Unlike in the previous case, it is easy to see that when $y$ goes to zero we have now $V(0)=\mp 1/(12 a_{0}^2)$, where it is important to take note of the sign inversion. This peculiarity leads to different dynamics according to the open or closed character of the Universe. For the closed case we can actually expand the potential in powers of $y$ in the same fashion we did before, to obtain
\begin{equation}
\dot{y}^2-\frac{\lambda}{12}y^2-\frac{a_{0}^{-2}}{12}\approx 0.\label{ecbajocerrado}
\end{equation}%
In this way the high energy regime for the scale factor is given by $a(t)\approx t$, with an associated Hubble rate $H\approx t^{-1}$. According to this picture the closed Universe possesses a singularity in $t=0$, and it expands for $t>0$ in an accelerated manner due to $\ddot{y}>0$.

\bigskip
A radically different picture emerges out when one considers the open Universe. Given that the potential goes to a positive number when $y\rightarrow 0$, and that it is negative when $y\rightarrow\infty$, it a has a root somewhere. Actually, by inspecting Eq. (\ref{potcaso2}) for the minus sign we easily see that the root results
\begin{equation}
y_{min}=\sqrt{k_{0}},\,\,\,\Rightarrow a_{min}=1/\sqrt{\lambda}.\label{ecbajocerrado}
\end{equation}%
Because the ``energy level'' in Eq. (\ref{ecpot}) is null, we observe that the Universe expands from the minimum size $a_{min}=\lambda^{-1/2}$. The functional form of the scale factor near this minimum can be obtained expanding the potential (\ref{potcaso2}) in the small quantity $y-y_{min}$. In this way we find that Eq. (\ref{ecpot}) reduces to
\begin{equation}
\dot{y}^2-\frac{\lambda \sqrt{k_{0}}}{6}\,y+\frac{\lambda \,k_{0}}{6}\approx0,\label{ecpotabiaprox}
\end{equation}%
leading to the scale factor
\begin{equation}
a(t)\approx a_{min}+\frac{\sqrt{\lambda}}{24}\,t^2.\label{facbouncing}
\end{equation}%
This constitutes a bounce of the scale factor in the event $t=0$, where $H=0$. Unlike the closed case considered above, the spacetime results now geodesically complete, and the cosmic evolution starts its accelerated expansion from a minimum volume given by $a_{min}^3=\lambda^{3/2}$ with a maximum energy density
\begin{equation}
\rho_{max}\propto a_{min}^{-4}=\lambda^{2}.\label{facbouncing}
\end{equation}%
This quantity can be interpreted as the maximum energy that can be stored as a consequence of the minimum volume existent due to the repulsive quantum effects governing the very early Universe. The reason why this occurs only in the context of open models remains unclear. This and many other open questions, such as the existence of regular, asymptotically flat vacuum black hole solutions, will be matter of future works.

\emph{Acknowledgments}. I would like to thank Rafael Ferraro for sharing so many years of enjoyable discussions and J. Areta for a careful reading of the manuscript. This work was supported by CONICET.

\end{document}